\documentclass[11pt, oneside]{article}   	
\usepackage{jheppub}

\usepackage{amsmath, amsthm, graphicx,amssymb,subfigure,bbm}
\usepackage[all]{xy}
\preprint{DIAS-STP-23-21\\
  \rightline{ QMUL-PH-23-33}}

\author[a]{Denjoe O'Connor,}
\affiliation[a]{School of Theoretical Physics, Dublin Institute for Advanced Studies, \\
       10 Burlington Road, Dublin 4, Ireland.}
\emailAdd{denjoe@stp.dias.ie}
\author[b,c]{Sanjaye Ramgoolam,}
\affiliation[b]{Centre for Theoretical Physics,  Department of Physics and Astronomy,\\
   Queen Mary University of London, 327 Mile End Road, London E1 4NS, UK.}
\emailAdd{s.ramgoolam@qmul.ac.uk}
\affiliation[b]{National Institute for Theoretical Physics,\\
  School of Physics and Centre for Theoretical Physics,\\
University of the Witwatersrand, Wits, 2050, South Africa}

\abstract{We give a path integral construction of the quantum
  mechanical partition function for gauged finite groups.  Our
  construction gives the quantization of a system of $d$, $N\times N$
  matrices invariant under the adjoint action of the symmetric group
  $S_N$.  The approach is general to any discrete group. For a system
  of harmonic oscillators, i.e. for the non-interacting case, the
  partition function is given by the Molien-Weyl formula times the
  zero-point energy contribution. We further generalise the result to a system
  of non-square and complex matrices transforming under arbitrary
  representations of the gauge group.}  \arxivnumber{}

\title{Gauged permutation invariant matrix quantum mechanics: Path Integrals}

\begin{document}
\maketitle
\section{Introduction}\label{intro}
Permutation invariant matrix models, defined by matrix integration
with a measure of the form $d^{N^2}{\kern -2.5pt}X e^{ -V(X) } $, where $X$ is an $N
\times N $ matrix of variables, $d^{N^2}{\kern-2.5pt}X$ is the standard Euclidean measure
on $N^2$ real variables and $V(X)$ is a potential invariant under
permutation transformations, were introduced in
\cite{Kartsaklis:2017aaa} with motivations from computational
linguistics. The general Gaussian potential $V(X)$, of polynomial degree
two in the matrix variables, invariant under permutation
transformations $ X \rightarrow U X U^T$ for $N \times N $ permutation
matrices $U$, was solved using the representation theory of the
symmetric group $S_N$ \cite{Ramgoolam:2018xty}. Under this
transformation by $S_N$ matrices, $X$ transforms as $V_N \otimes V_N$
where $V_N$ is the natural representation of $S_N$. The potential $V(X)$
contains two linear invariants and eleven quadratic ones, which were
each weighted by an independent parameter to give a solvable
$13$-parameter model.  The $13$-parameter models were used to find
evidence of approximate Gaussianity in ensembles of word matrices constructed from
natural language data \cite{RSS,ACRS}.  Motivated by holographic
properties of analogous gauged $U(N)$ invariant matrix theories
\cite{Malda,BBNS,CJR}, a large $N$-factorisation property for
permutation invariant observables in an inner product defined using
the simplest matrix model was established in
\cite{Barnes:2021tjp}.

Permutation invariant quantum mechanical matrix
models were described in \cite{Barnes:2022qli,Padellaro:2023blb}. An orthogonal basis of
states for the permutation invariant sector of the Hilbert space was
constructed for the simplest matrix harmonic oscillator, by employing the
representation theory of partition algebras. The canonical
quantization of the general $11$-parameter matrix harmonic oscillator,
defined using the $11$ quadratic permutation invariant functions of
$N\times N$ matrices, was given and the canonical partition function
was computed.

In this paper we realise the permutation transformation of the quantum
mechanical matrix degrees of freedom as a gauge symmetry within a path
integral formulation. We do this by beginning with a lattice
formulation of the canonical partition function. We take a lattice
with $\Lambda$ sites wrapped on a circle with period $\beta$ so that
the lattice spacing is $a=\beta/\Lambda$. Then a permutation invariant
potential is placed at each site with the different sites connected in
the standard lattice fashion by a discretized Laplacian but with the
hopping terms connected by permutation matrices to parallel transport
the permutation symmetry. This in effect is a lattice gauged matrix
model with the discrete gauge group $S_N$. In the context of lattice
gauge theory finite groups have in fact received significant attention
since the inception of the subject
\cite{Rebbi:1979sg,Creutz:1983ev,Drouffe:1983fv,Kogut:1982ds,Hasenfratz:1985pd,Hasenfratz:2000sa,Hasenfratz:2000hd}.
See \cite{Catterall:2022wjq} for a recent discussion.

The invariance of the partition function is enforced by summing over
the $\Lambda$ linking gauge variables which are representations of the
symmetric group in the natural representation acting by adjoint
action.  We first establish that all of the individual $\Lambda$ group
elements can be collected to the final connecting link as a single
group element using the invariance of the measure. The sum over gauge
fields then involves a single sum over the group elements. We then
concentrate on the Gaussian case where the matrices can be integrated out. This gives
an inverse square root of the determinant of the quadratic form in the lattice
action which in turn depends on the single group element.  The
lattice partition function is then obtained by summing over all group
elements. Finally we show that the resulting expression when the
zero-point energy is removed (i.e. when the ground state energy is
subtracted) is the classical Molien-Weyl expression (see \cite{TextBookMW} for a textbook discussion)
for the generating function of the Hilbert series of the quantum matrix model. This
allows us to easily discuss quantum models more generally, by
e.g. adding more matrices transforming under the group, or extending
to complex and general representations.

Our approach works quite generally for both continuous and discrete
groups and we establish the rather general result
(\ref{generalResult}) which is valid for a system of complex matrices
with Gaussian action. Furthermore, one can easily incorporate an
invariant local time-independent potential.

The principal results presented here are:
\begin{itemize}
\item We give a path integral derivation of the Molien-Weyl formula
  which in this context describes a system of non-interacting quantum
  matrix models with a $U(N)$ gauge invariance in the adjoint
  representation and in a thermal bath.
\item We then extend the derivation to include discrete groups such as
  the symmetric group $S_N$.  Modifying the potential as in
  \cite{Kartsaklis:2017aaa,Ramgoolam:2018xty,Barnes:2022qli,Padellaro:2023blb}
  allows us to access the 11-parameter family of $S_N$ invariant
  models. 
\item Finally we discuss the case when the gauge group acts in
  different representations on the left and right of the matrices
  which are now not necessarily square matrices. The analysis can
  easily be extended to gauged quiver matrix models.
\end{itemize}

The paper is organised as follows: To establish notation and introduce
the lattice formulation that we will use throughout we begin in
Section \ref{sec2} with a brief review of the path integral formulation
of a simple harmonic oscillator. In Section \ref{sec3} we then
describe a lattice version of the gauged system and show that on the
lattice, the local gauge transformations can be accumulated into a
holonomy around the thermal circle. This holonomy can be placed on any
link but we choose to place it on the closing link of the periodic
lattice. Section \ref{sec4} discusses the case of discrete groups and
recovers the classical Molien formula. Then in \ref{sec5} we extend the
discussion to arbitrary representations of the gauge groups. We end
the paper with some conclusions and discussion. The paper also
contains two short appendices where the relevant set of lattice
determinants arising in the bulk of the paper are evaluated explicitly.

\section{Path integral for a system of harmonic oscillator}\label{sec2}
A simple quantum oscillator is described by the Hamiltonian
$H=m(a^\dag a+\frac{1}{2})$. The partition function in this case is
\begin{equation}
Z(x)=\sum_{n=0}^{\infty}{\bf Tr}({\rm e}^{-\beta H})=\frac{x^{1/2}}{1-x}
\quad \hbox{where}\quad x={\rm e}^{-\beta  m}\, .
\end{equation}
Here we give a Euclidean lattice path integral derivation of this result
which we will then generalise to the Molien-Weyl formula.

The Euclidean action is obtained by taking the Lagrangian of the
oscillator, rotating to imaginary time on a circle of period
$\beta$. So for the oscillator the Euclidean action is then
\begin{equation}
  S[X]=\int_0^\beta d\tau \left(\frac{1}{2}(\partial_\tau X)^2+\frac{1}{2}m^2 X^2\right)
\end{equation}
The Euclidean lattice
version of this consists of replacing the time interval with a
periodic lattice of $\Lambda$ points so that $\tau\rightarrow\tau_n=n
a$, with $a=\beta/\Lambda$ the lattice spacing, and $n$ labeling the
lattice points from $1$ to $\Lambda$. Then we denote $X(\tau_n)=X_n$
with periodicity implying $X_{\Lambda+1}=X_1$. The Euclidean time
derivative is then replaced by a finite difference which we take to be
\begin{equation}
  \partial_\tau X\rightarrow \frac{X_{n+1}-X_n}{a}=\frac{{\rm e}^{a\partial_\tau}-1}{a}X_n\, ,\quad d\tau\rightarrow a
\label{Partial_Lat}
\end{equation}
and where ${\rm e}^{a\partial_\tau}$ is the shift operator on the lattice, taking $X_n$ to $X_{n+1}$
The lattice action can then be expressed as
\begin{equation}
    \label{LatAction1}
  S_\Lambda[X]=\sum_{n,n'=1}^\Lambda \frac{a}{2}X_{n'}(\Delta_\Lambda+m^2)_{n',n}X_n\, ,
\end{equation}
where $\Delta_\Lambda$ is the lattice version of $-\partial_\tau^2$ and the only non-zero elements of $(\Delta_\Lambda+m^2)_{n',n}$ are
\begin{equation}
  (\Delta_\Lambda+m^2)_{n,n}=\frac{2}{a^2}+m^2\qquad (\Delta_\Lambda+m^2)_{n+1,n}=(\Delta_\Lambda+m^2)_{n,n+1}=-\frac{1}{a^2}\, .
  \end{equation}
The path integral is then given by the $\Lambda$ dimensional Gaussian integration over $X_n$ and yields
\begin{equation}\label{ZLatSHO}
  Z_\Lambda=\int_{{\mathbb R}^\Lambda} \frac{d^\Lambda X}{(2\pi a)^{\Lambda/2}}{\rm e}^{-S_\Lambda[X]}
  ={\rm Det}^{-1/2}(a^2(\Delta_\Lambda+ m^2)) 
\end{equation}
where
\begin{equation}\label{ZDetSHO}
  \Delta_\Lambda=\frac{2-{\rm e}^{a\partial_\tau}-{\rm e}^{-a\partial_\tau}}{a^2}\, .
\end{equation}
The matrix
\begin{equation}\label{M_Lambda_1}
a^2(\Delta_\Lambda+m^2)=2+\frac{\beta^2m^2}{\Lambda^2}-{\rm e}^{a\partial_\tau}-{\rm e}^{-a\partial_\tau}:=M_{\Lambda,1}
\end{equation}
is a tri-diagonal matrix with $2+\frac{\beta^2m^2}{\Lambda^2}$ on the diagonal, $-1$ in the leading off diagonals and the upper right and lower left corners (see (\ref{MLambda}) and the subsequent discussion).

The $\Lambda$ eigenvectors of $a^2\Delta_\Lambda$, are obtained by noting
the discrete lattice translation invariance implies that translation by $\Lambda$ steps (i.e. corresponding to translations around
one full period, $\beta=\Lambda a$)  gives the identity i.e. ${\rm e}^{\Lambda a\partial_\tau}=1$ and the operator is diagonalized by a discrete Fourier transform. The eigenvalues are then given in terms of the roots of unity of ${(w_k)}^{\Lambda}=1$, i.e. $w_k={\rm e}^{2\pi i k/\Lambda}$. The corresponding eigenvectors are $v_n^k=\frac{1}{\sqrt{\Lambda}}{\rm e}^{\frac{2\pi i n k}{\Lambda}}$ with $n$ the lattice site. Colectively $v_n^k$ form the unitary matrix
that diagonalizes $\Delta_\Lambda$ and $M_{\Lambda,1}$.  The eigenvalues of
$M_{\Lambda,1}$ are then
\begin{equation}
  \lambda_k=2-2\cos(\frac{2\pi k}{\Lambda})+\frac{\beta^2 m^2}{\Lambda^2}
  \label{LatLapspectrum}
\end{equation}
so that the partition function (\ref{ZLatSHO}) is
\begin{equation}
  Z_\Lambda={\rm Det}^{-1/2}M_{\Lambda,1}=\prod_{k=1}^\Lambda \lambda_k^{-1/2}.
\end{equation}
One can now compute $Z_\Lambda$ by factoring the $\lambda_n$, i.e.  
writing $2\cos(\frac{2\pi k}{\Lambda})=w_k+1/w_k$, factoring
\begin{equation}
  \lambda_k=2+\mu^2-w_k-1/w_k=\frac{1}{w_k}(z_+-w_k)(w_k-z_-)
  \end{equation}
and using
\begin{equation}
\prod_{k=1}^{\Lambda}(z-w_k)=z^\Lambda-1\quad \hbox{and}\quad \prod_{k=1}^\Lambda(-w_k)=-1
\end{equation}
to find
\begin{equation}\label{DetMLamda1Eval}
 {\rm Det}M_{\Lambda,1}= \prod_{k=1}^\Lambda \lambda_k=(z_{+}^\Lambda-1)(1-z_{-}^{\Lambda})=z_{+}^\Lambda+z_{-}^\Lambda-2
\end{equation}
so that the lattice partition function is given by
\begin{equation}\label{ZLambda}
  Z_\Lambda=(z_+^\Lambda -1)^{-\frac{1}{2}}(1-z_-^\Lambda)^{-\frac{1}{2}}=
    \frac{z_-^{\frac{\Lambda}{2}}}{1-z_-^\Lambda}
\end{equation}
where $z_{\pm}=1+\frac{\mu^2}{2}\pm \sqrt{\mu^2(1+\frac{\mu^2}{4})}$,
$\mu=\frac{m\beta}{\Lambda}$ and $z_{+}z_{-}=1$.

The continuum limit is obtained by noting that
$\displaystyle\lim_{\Lambda\rightarrow\infty}z_{-}^\Lambda\rightarrow {\rm e}^{-\beta m}=x$ so that
\begin{equation}
Z_\Lambda=\frac{z_{-}^\frac{\Lambda}{2}}{1-z_-^\Lambda}\longrightarrow\displaystyle\lim_{\Lambda\rightarrow\infty} Z_\Lambda=\frac{{\rm e}^{-\frac{m\beta}{2}}}{1-{\rm e}^{-m\beta}} =\frac{x^{1/2}}{1-x}.
\end{equation}
as required.  Hence, by simply taking the limit of an infinite number of lattice
points with fixed $\beta m$ one recovers exactly the continuum result.
\section{Gauged Gaussian Matrix Models.}\label{sec3}
We now consider $X$ to be an $N \times N$  hermitian matrix and the action
involves a trace over matrices. The resulting action has a global
$U(N)$ invariance\footnote{This is effectively a global $SU(N)$
invariance since the $U(1)$ cancels in the adjoint action. One could
equally consider traceless matrices, but it convenient to leave the
$U(1)$ and the trace in our expressions.} under the adjoint action of
$U(N)$ so that $X\rightarrow g X g^{{-1}}$. This invariance can be
promoted to a local gauge symmetry. The Euclidean action is then
\begin{equation}\label{ContinuumGaugeAction}
  S[X]=\int_0^\beta d\tau {\bf tr}\left(\frac{1}{2}({\cal D}_\tau X)^2+\frac{1}{2}m^2 X^2\right)\end{equation} 
where ${\cal D}_\tau=\partial_\tau-i[A,\,\cdot\,]$ is the covariant derivative with
the gauge field $A(\tau)$ being an $N\times N$ hermitian matrix.

When discretising the covariant derivative ${\cal D}_\tau$ we need the lattice analogue of (\ref{Partial_Lat}). It is naturally defined to be
\begin{equation}
  \label{LatticeCovDeriv}
  {\cal D}_\tau X\rightarrow \frac{g_{n,n+1}X_{n+1}g_{n+1,n} -X_n}{a}\ =\frac{{\rm e}^{a{\cal D}_\tau}-1}{a^2}X_n
\end{equation}
where $g_{n,n+1}$ are the link gauge variables that parallel transport the group element at $n+1$ back to that at $n$ and $g_{n+1,n}=g_{n,n+1}^{-1}$.
The corresponding lattice Laplacian is
given by 
\begin{equation}
  \label{LatticeCovLap}
  \Delta_{\Lambda,g}=\frac{2-e^{a {\cal D}_\tau}-e^{-a {\cal D}_\tau}}{a^2}\
  \end{equation}
and the action is a quadratic form as in (\ref{LatAction1}) with $\Delta_{\Lambda}$ replaced by $\Delta_{\Lambda,g}$. By substituting a path ordered product, denoted  ${\cal P}$, one sees 
\begin{equation}\label{linkU}
g_{n,n+1} ={\cal P}\exp\left[i\int_{na}^{na+a}d\tau\,A(\tau)\right] \ ,
\end{equation}
and taking $a$ to zero in (\ref{LatticeCovLap}) one recovers the continuum Laplacian.
The corresponding partition function is then
\begin{equation}\label{ZNLambda_with-gs}
  Z_{N,\Lambda}=\int_{{U(N)}^\Lambda} \int_{{\mathbb R}^{N^2\Lambda}}\prod_{k=1}^\Lambda\mu(g_{k,k+1})\frac{d^{N^2}X_k}{(2\pi a)^{N^2}}{\rm e}^{-S_{\Lambda,g}}
\end{equation}
where
\begin{equation}\label{GaugeLatAction_v1}
S_{\Lambda,g}=\sum_{n,n'=1}^{\Lambda}\frac{a}{2}{\bf tr}(X_n' (\Delta_{\Lambda,g}+m^2)_{n',n}X_n)\, .
\end{equation}
Eqn (\ref{ZNLambda_with-gs}) is a matrix integral over the matrix variables, $X_n$, with Dyson measure $d^{N^2}{\kern -2.5pt}X_n$, for each lattice site,
$n$, together with $\Lambda$ integrals over the
individual link group elements $g_{n,n+1}$. These latter integrals have Haar measure, $\mu(g)$, normalised to unity.

As shown in \cite{Filev:2015hia} one can further simplify the dependence of the action  to a dependence on a single group element.
First observe that reorganising the lattice action (\ref{GaugeLatAction_v1}) gives
\begin{equation}\label{GaugeLatAction}
S_{\Lambda,g} = \sum_{n=1}^{\Lambda}\frac{1}{a}{\bf tr}\left\{-\,X_ng_{n,n+1}X_{n+1}g_{n+1,n}+(1+\frac{\beta^2 m^2}{2\Lambda^2}) X_n^2\right\}\ ,
\end{equation}
and we have used that periodicity of the lattice implies $X_{\Lambda+1}=X_1$.
Since the $X_n$ will be integrated over we can make a change of variables $X'_1=X_1'$,  $X'_2=g_{1,2}X_2 g_{2,1}$ so that $X_1g_{1,2}X_2 g_{2,1}$ becomes $X_1'X_2'$, and $X_2g_{2,3}X_3g_{3,2}=g_{2,1}X_2'g_{1,2}g_{2,3}X_3g_{3,2}$ which, with cyclicity under the trace, allows us to redefine $X_3'=g_{1,2}g_{2,3}X_3g_{3,2}g_{2,1}$, etc. We can continue, using the transformation 
\begin{eqnarray}\label{Redefining_X}
  &&{X'}_{r} = \prod_{k=1}^{r-1}g_{k,k+1}\,X_{r}\,\prod_{k=1}^{r-1}g_{k+1,k}\ ,
\end{eqnarray}
until we arrive at $X_\Lambda g_{\Lambda,1}X_1 g_{1,\Lambda}$ which becomes $X_\Lambda'g_{1,2}\dots g_{\Lambda,1}X_1'g_{1,\Lambda}\dots g_{2,1}$. Due to periocicity of the lattice the accumulated group element, $g=g_{1,2}\dots g_{\Lambda,1}=\prod_{k=1}^{\Lambda}g_{k,k+1}$, can no longer be absorbed\footnote{Note $g_{1,\Lambda}\dots g_{2,1}=g^{-1}$.}.
The group elements in $g$ form a closed loop, and the product is a holonomy element. This holonomy element, $g$, can be placed anywhere on the circle but here it is placed on the link between $n=\Lambda$ and $n=1$. The lattice action then becomes
\begin{equation}\label{SLambda_g}
S_{\Lambda,g}= -\frac{1}{a}{\bf tr}\left\{\sum_{n = 1}^{\Lambda-1}{X'}_n{X'}_{n+1}+{X'}_{\Lambda}g\,{X'}_1 g^{-1}\right\}+\frac{1}{a}\sum_{n=1}^{\Lambda-1}{\bf tr}\left\{(1+\frac{a^2\beta^2 m^2}{2}){X'}_n^2\right\}\ .
\end{equation}
The structure of (\ref{SLambda_g}) is now of the form (\ref{ZLatSHO})
\begin{equation}
  S_{\Lambda,g}=\sum_{n,n'=1}^{\Lambda}\frac{1}{2}{\bf tr}X_{n'} a(\Delta_{\Lambda,g}+m^2{\bf 1})_{n',n} X_n
\end{equation}
where ${\bf 1}$ is the unit matrix and group identity element.  The
matrix $(a^2\Delta_{\Lambda,g}+\frac{\beta^2m^2}{\Lambda^2}{\bf  1})_{n',n}$
is a $\Lambda N^2$ dimensional tri-diagonal matrix with
$g\otimes g^{-1}$ in the right upper corner and its inverse in the
lower corner. The diagonal elements are all
$(2+\frac{\beta^2  m^2}{\Lambda^2}){\bf 1}$ and the off diagonals are $-{\bf 1}$.

The group invariance and normalization to unity of the $\mu(g)$ measure ensures that the resulting path integral has a single
integration over this final gauge transformation, $g$.
The partition function (\ref{ZNLambda_with-gs}) is then
\begin{equation}\label{ZNLambda}
Z_{N,\Lambda}=\int \mu(g){\rm\bf Det}^{-1/2}(a^2\Delta_{\Lambda,g}+\frac{\beta^2 m^2}{\Lambda^2}{\bf 1}))=\int \mu(g){\rm\bf Det}^{-1/2}M_{\Lambda,g}
\end{equation}
where
\begin{equation}
 M_{\Lambda,g}:=a^2\Delta_{\Lambda,g}+\frac{\beta^2m^2}{\Lambda^2}=2+\mu^2-e^{a {\cal D}_\tau}-e^{-a {\cal D}_\tau}
  \end{equation}
and our notation is ${\rm\bf Det}$ for the $\Lambda N^2\times \Lambda N^2$
determinant, ${\rm Det}$ for the $\Lambda\times \Lambda$ lattice determinant and ${\rm\bf det}$ for the $N^2\times N^2$ or group determinant.

The group element entering the expressions creates no additional complication as it acts as a commuting variable with respect to the $\Lambda\times\Lambda$ matrix entries and as shown in the Appendix \ref{appendixA} one arrives at the determinant of the lattice determinant as 
\begin{equation}\label{ZNLambdaEigForm}
{\rm Det}M_{\Lambda,g}=z_{+}^\Lambda+z_{-}^\Lambda-g\otimes g^{-1}-g^{-1}\otimes g=z_{+}^\Lambda({\bf 1}-z_{-}^{\Lambda}g\otimes g^{-1})({\bf 1}-z_{-}^{\Lambda}g^{-1}\otimes g)\, .
\end{equation}
The partition function of the gauged system (\ref{ZNLambda})
  is then given by taking the remaining matrix determinant
  (which we denote ${\rm\bf det}$) and integration over the gauge group gives the lattice partition function
\begin{equation}\label{ZLambda_g}
  Z_{N,\Lambda}=\int \mu(g) {\rm\bf det}[(z_+^\Lambda{\bf 1} -g^{-1}\otimes g)({\bf 1}-z_-^\Lambda g\otimes g^{-1})]^{-\frac{1}{2}}
  = \int \mu(g)\frac{z_{-}^{\frac{N^2\Lambda}{2}}}{{\rm\bf det}[{\bf 1}-z_-^\Lambda g\otimes g^{-1}]}.
\end{equation}
To obtain this we have used that $z_{-}=z_{+}^{-1}$ and that the adjoint representation is Hermitian so that
${\rm\bf det}[{\bf 1}-z_-^\Lambda g\otimes g^{-1}]={\rm\bf det}[{\bf 1}-z_-^\Lambda g^{-1}\otimes g]$.
    
We can now take the limit of $\Lambda$ to infinity keeping $\beta m$ fixed to obtain the continuum result
\begin{equation}
  Z_{N,\infty}= \int \mu(g)\frac{{\rm e}^{-\frac{N^2\beta m}{2}}}{{\rm\bf det}[{\bf 1}-{\rm e}^{-\beta m} g\otimes g^{-1}]}\, .
\end{equation}
The numerator ${\rm e}^{-\frac{N^2\beta m}{2}}$ comes from the zero
point energy of the quantum system of oscillators.  If we remove this
zero-point energy to consider the normal ordered Hamiltonian we obtain
the generating function for the Hilbert-Poincar\'e series as the Molyen-Weyl formula
\begin{equation}
Z_N(x)=\int \mu(g)\frac{1}{{\rm\bf det}[{\bf 1}-x g\otimes g^{-1}]}
\end{equation}
For the gauge group $U(N)$ we can simplify this expression further by diagonalizing the holonomy matrix $g$. Diagonalization reduces it to
the Cartan torus with eigenvalues $z_i={\rm e}^{i\theta_i}$. The measure becomes a product of Vandermonde determinants
\begin{equation}
  \Delta(z)={\displaystyle\prod_{1\leq i < j\leq N}}(z_i-z_j)
\end{equation}
 and we have the explicit expression
\begin{equation}
  Z_N(x)=\frac{1}{N!}\oint \prod_{i=1}^N\frac{dz_i}{2\pi i}\Delta(z)\Delta(z^{-1}) \prod_{i=1}^N\prod_{j=1}^N\frac{1}{1-x z_i z_j^{-1}}
\end{equation}
The $N$ contour integrals are around the unit circle and can be evaluated in sequence to give
\begin{equation}
Z_N(x)=\prod_{n=1}^N\frac{1}{1-x^n}=P_N(x)=\sum_{n=0}^\infty p_N(n)x^n
\end{equation}
i.e. $Z_n(x)=P_N(x)$ is the generating function of partitions of $n$ with at most $N$ parts.

For a system of $d$ matrices transforming simultaneously as $X_a\rightarrow g X_a g^{-1}$ with all of the masses $m_a$ 
distinct, so that $x_a={\rm e}^{-\beta m_a}$, the resulting normal ordered partition function is
\begin{equation}\label{Molien-Weyl-d-matrices}
  Z_N(x_1,\cdots,x_d)=\frac{1}{N!}\oint \prod_{i=1}^N\frac{dz_i}{2\pi i}\Delta(z)\Delta(z^{-1}) \prod_{a=1}^d\prod_{i=1}^N\prod_{j=1}^N\frac{1}{1-x_a z_i z_j^{-1}}
\end{equation}
The case of $d=2$ has received much attention due to its relevance to the counting of $\frac{1}{4}$-BPS states in ${\cal N}=4$ supersymmetric Yang-Mills \cite{Benvenuti:2006qr,Dolan:2007rq}.

The specialisation to $SU(N)$ from $U(N)$ is rather trivial since in the path integral formulation the additional degree of freedom associated
under the adjoint action is a $U(1)$ factor which cancels. Similarly the trace of each matrix is itself a singlet and removing the trace is
equivalent to multiplying $Z_n(x_1,\cdots,x_d)$ by $\prod_{a=1}^d(1-x_a)$.

\section{Discrete Groups and More General Representations}\label{sec4}
Taking the gauge group to be discrete causes little
complication. Though it is not clear how to think of the continuum
expression (\ref{ContinuumGaugeAction}) with $X(\tau)\rightarrow
g(\tau)X(\tau)g(\tau)^{-1}$ for a $g(\tau)\in G$ with $G$ a discrete
group, such as the symmetric group $S_N$, the lattice expressions
(\ref{LatticeCovDeriv}) and (\ref{LatticeCovLap}) and lattice actions
(\ref{GaugeLatAction_v1}) and (\ref{GaugeLatAction}) and the steps
leading to (\ref{SLambda_g}) (where we collect the gauge
transformations on the final link) make perfect sense. We can
therefore proceed as in the continuous group case but with the Haar measure
replaced by $1/\vert G\vert$ and the integration by the sum over the
group elements.

The sum of lattice determinants (\ref{ZNLambda}) in the finite group
case then becomes
\begin{equation}\label{ZNLambda-finiteG}
  Z_{N,\Lambda}=\frac{1}{\vert G\vert}\sum_{g\in G}{\rm\bf Det}^{-1/2}(a^2(\Delta_{\Lambda,g}+\beta^2 m^2{\bf 1}))=\frac{1}{\vert G\vert}\sum_{g\in G}{\rm\bf Det}^{-1/2}M_{\Lambda,g}
  \end{equation}
We proceed (see the appendix \ref{appendixA}) by expanding the determinant in minors  and
establish
\begin{equation}\label{LatDet-fintieG}
{\rm Det}M_{\Lambda,g}=z_{+}^\Lambda+z_{-}^{\Lambda}+g\otimes g^{-1}-g^{-1}\otimes g\, .
\end{equation}
directly.

From (\ref{LatDet-fintieG}) by taking $\Lambda$ to infinity in (\ref{ZNLambda-finiteG}) we obtain the classical Molien form of the expression as
\begin{equation}\label{ClassicalMolien}
Z_N(x)=\frac{1}{\vert G\vert}\sum_{g\in G}\frac{1}{{\rm\bf det}[{\bf 1}-x g\otimes g^{-1}]}
\end{equation}
In the case of $G=S_N$, the symmetric group on $N$ elements, 
explicitly for $N=2,3$ and $4$ we have 
\begin{eqnarray}
  &\hspace{-4pt}
  Z_{S_2}(x)&=\frac{1}{2!}\frac{1}{(1-x)^4}+\frac{1}{2}\frac{1}{(1-x^2)^2}\nonumber\\
  &\hspace{-4pt}
  Z_{S_3}(x)&=\frac{1}{3!}\frac{1}{(1-x)^9}+\frac{1}{2}\frac{1}{(1-x)(1-x^2)^4}+\frac{1}{3}\frac{1}{(1-x^3)^3}\\
  &\hspace{-4pt}
  Z_{S_4}(x)&=\frac{1}{4!}\frac{1}{(1-x)^{16}}+\frac{1}{4}\frac{1}{(1-x)^4(1-x^2)^6}+\frac{1}{2!2^2}\frac{1}{(1-x^2)^8}+\frac{1}{3}\frac{1}{(1-x)(1-x^3)^5}
  +\frac{1}{4}\frac{1}{(1-x^4)^4}\,.\nonumber
  \end{eqnarray}
We have written the expression with the individual symmetry factors and used that for $N=2$ the two elements of $S_2$ in the natural representation
can be taken to be the identity and the Pauli matrix $\sigma^1$. For $N=3$ the sum (\ref{ClassicalMolien}) is over the 6 possible permutations
on three elements
\begin{equation}\label{S3matrices}
\{\left(\begin{array}{ccc}
 1 & 0 & 0 \\
 0 & 1 & 0 \\
 0 & 0 & 1 \\
\end{array}
\right),\quad
\left(
\begin{array}{ccc}
 0 & 1 & 0 \\
 1 & 0 & 0 \\
 0 & 0 & 1 \\
\end{array}
\right),\quad
\left(
\begin{array}{ccc}
 0 & 0 & 1 \\
 0 & 1 & 0 \\
 1 & 0 & 0 \\
\end{array}
\right),\quad
\left(
\begin{array}{ccc}
 1 & 0 & 0 \\
 0 & 0 & 1 \\
 0 & 1 & 0 \\
\end{array}
\right),\quad
\left(
\begin{array}{ccc}
 0 & 1 & 0 \\
 0 & 0 & 1 \\
 1 & 0 & 0 \\
\end{array}
\right),\quad
\left(
\begin{array}{ccc}
 0 & 0 & 1 \\
 1 & 0 & 0 \\
 0 & 1 & 0 \\
\end{array}\right)
\}\, 
\end{equation}
with the identity matrix giving the first term, the next 3 matrices in (\ref{S3matrices}) giving the second and the last two giving the final term.

The sum of group elements breaks into conjugacy classes of permutations with the permutations labeled by partitions of $N$.  So
\begin{equation}\label{SumOfPartitionsSmallN_v2}
  Z_{S_N}(x)=\frac{1}{N!}\sum_{p\vdash N}\sum_{g\in C(p)}\frac{1}{{\rm\bf det}[{\bf 1}-x g\otimes g^{-1}]}=\sum_{p\vdash N}\frac{1}{{\rm Sym}_p}\frac{1}{{\rm\bf det}[{\bf 1}-x g_p\otimes g_p^{-1}]}.
\end{equation}
where $C(p)$ is the conjugacy class of $p$, ${\rm Sym}_p$ is the number of elements of the stabiliser group of the conjugacy
class (so that $\frac{N!}{{\rm Sym}_p}$  is the number of elements in $C(p)$) and $g_p$ is a single representative of the conjugacy class.
The symmetry factor ${\rm Sym}_p$ is given by  
\begin{equation}
{\rm Sym}_p = \prod_{i=1}^s a_i^{p_i}p_i! 
\end{equation}
where we have used the form $p =[a_1^{p_1},a_2^{p_2},\cdots, a_{s}^{p_s}]$, with 
$a_i$ distinct non-zero parts (cycle lengths) such that $1\le a_i \le N $ and $p_i$ are positive integers ordered as $a_1 < a_2 < \cdots < a_s$ and 
$N=\sum_{i =1}^s a_i p_i$,

By choosing the representative one can evaluate the determinant explicitly \cite{OConnorRamgoolam-II} and show that
in we show that
\begin{equation}\label{MainProp}  
Z_{S_N}(p,x)=\frac{1}{{\rm\bf det}[{\bf 1}-x g_p\otimes g_p^{-1}]}=\prod_{i}{1\over(1-x^{a_i})^{a_i p_i^2}}
 \prod_{i<j}{1\over(1-x^{L(a_i,a_j)})^{2G(a_i,a_j)p_ip_j}}
\end{equation}
where $ L ( a_i , a_j ) $ is the least common multiple of $ a_i $ and $ a_j $ and $G ( a_i , a_j ) $ is the
greatest common denominator of $a_i , a_j$. 

By including the full generality of the quadratic invariant potential $V(x)=G_{ijkl}X_{ij}X_{kl}$, with $G$ in general involving 11-parameters and discussed in \cite{Ramgoolam:2018xty,Barnes:2022qli}, we can access
the full 11-parameter (7 independent variables when diagonalized) $S_N$ invariant partition functions. The resulting expression
\begin{equation}
  Z_N(G)  =\sum_{p\vdash N}\frac{1}{{\rm Sym}_p}\frac{1}{{\rm\bf det}[{\bf 1}-G g_p\otimes g_p^{-1}]}
  \end{equation}
will factor into a product form as described in \cite{OConnorRamgoolam-II}.

\section{Generalizations to more general representations}\label{sec5}
One may further generalize these arguments to cover the case when the matrix $X$ is neither Hermitian nor square and has a different representation acting
on the left and right so that the group action is  $X\rightarrow g_L(g)Xg_R^{-1}(g)$. The Euclidean action for an $N_L\times N_R$ matrix then becomes
\begin{equation}
  S[X]=\int_0^\beta d\tau \left({\bf tr} ({\cal D}_\tau X^\dag)({\cal D}_\tau X)+m^2 X^\dag X\right)
\end{equation} 
The arguments for accumulating the gauge transformations on the final link still go through and the lattice version becomes
\begin{equation}
  S_{\Lambda,R}=\sum_{n,n'=1}^{\Lambda}a{\bf tr}X_n^\dag (\Delta_{\Lambda,R}+m^2{\bf 1})_{n,n'} X_{n'}
  \end{equation}
where $R$ is the accumulated group action on the left and right, i.e. $R(g)=g_{L}(g)\otimes g_R^{-1}(g)$.
The integration over $X$ and $X^\dag$ now yields\footnote{Here the comlex Gaussian integration
  $$\int\frac{dzd\overline{z}}{\pi a }{\rm e}^{-M z\overline{z}}=\int \frac{dx dy}{2\pi a}{\rm e}^{-\frac{1}{2}M(x^2+y^2)}=\frac{1}{a M}\, .$$}
\begin{eqnarray}
  Z_{\Lambda,R}&=&{\rm\bf Det}^{-1}[a^2\Delta_{\Lambda,R}+\frac{\beta^2m^2}{\Lambda^2}{\bf 1}]\nonumber\\
  &=&z_{+}^{-N_LN_R\Lambda}{\rm\bf det}^{-1}\left[({\bf 1}-z_{-}^\Lambda R(g))({\bf 1}-z_{-}^{\Lambda}R(g)^{-1})\right]
\end{eqnarray}
The factor, $z_{+}^{-N_LN_R\Lambda}$  in the $\Lambda\rightarrow\infty$ limit gives $z_{+}^{-N_LN_R\Lambda}\rightarrow {\rm e}^{-N_LN_R\beta m}$ which is the contribution from the zero-point energy.
In this limit the normal-ordered expression corresponding the Molien-Weyl formula is therefore given by
\begin{equation}\label{generalComplex}
  Z_{N_L,N_R}(x)=\frac{1}{\vert G\vert}\sum_{g\in G}\frac{1}{{\rm\bf det}[(1-x R(g))(1-x R^{-1}(g))]}
  \end{equation}
or correspondingly an integral over the Haar measure in the case of a continuous group.

One can equally take a family of non-interacting matrices that collectively transform under the group i.e. $X^i\rightarrow g_L(g)X^ig_R(g)^{-1}$, $i=1,\cdots,d$ the resulting integrations over the matrices $X^A$ lead to determinants and
the evaluations proceed as before, so that we obtain
\begin{equation}\label{generalResult}
  Z_{N_L,N_R}(x,d)=\frac{1}{\vert G\vert}\sum_{g\in G}\frac{1}{{\rm\bf det}^d[(1-x R(g))(1-x R^{-1}(g))]}
  \end{equation}
of the equivalent expression with Haar measure for a continuous group.

The case of fermions is treated in the companion paper
\cite{OConnorRamgoolam-II} where the partition function is
derived\footnote{The lattice derivation is not given here as it
involves the additional complication of adding a Wilson term to deal
with Fermion doubling and it would take us outside the scope of this
paper.}. For a system of $d$ matrices of fermionic oscillators
transforming in the adjoint representation it is given by
\begin{equation}\label{Fermions}
  Z^{Fermions}_{N}(x,d)=\frac{1}{\vert G\vert}\sum_{g\in G}{\rm\bf det}^d[1+x g\otimes g^{-1}]
  \end{equation}
of the equivalent expression with Haar measure for a continuous group.

A further possible generalization is to restrict the group elements, $g_{n,n+1}$ in (\ref{GaugeLatAction}),
to a given conjugacy class $C(p)$. The resulting path integral will again give
the Molien-Weyl expression
\begin{equation}\label{ConjugacyClassZ}
  Z_N(C(g)) =\sum_{g\in C(g)}\frac{1}{{\rm\bf det}[{\bf 1}-G g\otimes g_p^{-1}]}=\frac{N!}{{\rm Sym}_p}Z_{S_N}(p,x).
\end{equation}
The normalization $\frac{N!}{{\rm Sym}_p}$ accounts for the number of elements in the conjugacy class.
Eqn (\ref{ConjugacyClassZ}) then provides a path integral interpretation for the individual $p$ dependent partition functions $Z_{S_N}(p,x)$.
Equivalently, they correspond to path integrals with holonomy around the thermal circle constrained to a fixed conjugacy class.

\section{Conclusions}\label{secConc}

In the above we provided a path integral derivation for the partition
function for a gauged matrix quantum mechanical system.  We
concentrated on the Gaussian case corresponding to a system of
oscillators subject to the constraint that all states are invariant
under the group action.  We began with a review of the case of a continuous $U(N)$
symmetry and the evaluation of the partition function led to the
Molien-Weyl formula for counting $U(N)$ invariant states of a system of oscillators
(\ref{Molien-Weyl-d-matrices}) when the zero
point energy was removed\footnote{The zero-point energy is unaffected by gauging the
system as the ground state is a gauge singlet.}.
The two matrix version of the $U(N)$
invariant case arises in ${\cal N}=4$ super Yang-Mills as the counting
problem for $\frac{1}{4}$-BPS states \cite{Dolan:2007rq}.

We then proceeded along similar lines for the finite group case beginning
again with a lattice formulation where the gauge field gives parallel
transport on the links, replacing the continuous group elements with
finite group elements. Our main interest was in the permutation group
$S_N$, however, the analysis is equally applicable to any finite
group. Taking the continuum limit led us to the Molien formula (\ref{ClassicalMolien})
for the counting of states in the finite group case. Using this for $S_N$
gives a sum over partitions  \eqref{SumOfPartitionsSmallN_v2} which can be analysed directly
for small $N$ and leads to our general result (\ref{MainProp}) which is derived and 
described in detail in \cite{OConnorRamgoolam-II}.

We were able to extend the formulation to arbitrary representations
and obtain a form for the general complex case which involves the
product of two Molien-Weyl factors (see eq (\ref{generalComplex})).

The extension to Fermions is straightforward but a lattice formulation
requires the elimination of fermion doublers. This is largely a
technical complication and we have not gone into these details here. In the
companion paper \cite{OConnorRamgoolam-II} it is shown that the final
result is as expected. One gets a determinant instead of an inverse
determinant, and the variable $x$ is replaced by $-x$ due to the
fermions requiring anti-periodic thermal boundary conditions.

It would also be interesting to study $S_N$ invariant systems in
higher dimensions. These can help elucidate features of higher
dimensional non-abelian theories along the lines of
\cite{Rebbi:1979sg} where the finite subgroups of $SU(2)$ were studied
on the lattice. A significant difference between the $U(N)$ and $S_N$
invariant quantum matrix models is the absence of the Vandermonde
determinant in the case of $S_N$. For $U(N)$ gauge systems the
Vandermonde is responsible for repulsion between eigenvalues of the
$U(N)$. This sets up a competition between the thermodynamic and
entropic aspects of the model and as the temperature is reduced leads
to a Hagedorn transition.  The absence of this repulsion in the $S_N$
and more generally for finite groups leads one to expect the
corresponding Hagedorn transition to be driven to zero
temperature. This is presumably why the critical temperature was
driven to zero in the old study of Rebbi \cite{Rebbi:1979sg}.  We will
return to the thermodynamics of the permutation invariant matrix
quantum mechanical systems in a future paper
\cite{OConnorRamgoolam-III}.

Including interactions such as a general potential $V(X)$ for the
matrices causes no additional difficulties other than the usual need
to resort to perturbation theory. The determinants that arise in the
current analysis would be the 1-loop approximations to the partition
functions of the interacting systems and one could proceed
perturbatively to evaluate approximate partition functions. It is
clear, however, from our discussion that there are no additional
difficulties introduced by including interactions for the discrete
gauged system and our lattice formulation would provide the starting
point for non-perturbative Hybrid Monte Carlo studies. It would be
especially interesting to find novel $S_N$ invariant interacting
systems such as supersymmetric interacting models or potential
generalisations of the BFSS  \cite{bfss}  and BMN \cite{bmn} matrix models of M-theory. These latter models
have received intense interest and study by hybrid Monte Carlo methods in recent years
see for example \cite{Filev:2015hia,Berkowitz:2016jlq} for the BFSS model and 
\cite{Catterall:2010gf,Asano:2018nol,Pateloudis:2022oos} for the BMN model.
\vskip.3cm

\centerline{\bf{Acknowledgments}}
\vskip.2cm 

DO'C is supported by the Irish Research Council and Science Foundation
Ireland under grant SFI-IRC-21/PATH-S/9391.  SR is supported by the
Science and Technology Facilities Council (STFC) Consolidated Grants
ST/P000754/1 “String theory, gauge theory and duality” and
ST/T000686/1 “Amplitudes, strings and duality”. SR thanks the Dublin
Institute for Advanced Studies for hospitality while part of this work
was being done.  We thank Yuhma Asano, George Barnes, Samuel Kov\'a\c{c}ik,
Veselin Filev and  Adrian Padellaro for discussions related to
the subject of this paper.

\begin{appendix}
\section{Evaluating Lattice Determinants}\label{appendixA}

Here we evaluate ${\rm\bf Det}({a^2\Delta_{\Lambda,g}+\mu^2})$ where all the gauge field is accumulated in the final link can be evaluated directly by noting 
 that it is a tri-diagonal matrix with the gauge field in the right upper corner and its inverse in the lower corner. The diagonal elements are
  all $2+\mu^2$ and the off diagonals are $-1$ and perform a determinant expansion by minors. 
  
 Let us evaluate the determinant of the more general matrix 
\begin{eqnarray}\label{MLambda}
  M_{\Lambda}=                                                                                                                                                   
    {\scriptsize                                                                                                                                                 
      \left(\begin{array}{cccccccc}
      d_1 & -\alpha_1 & 0 & 0 & 0 & \cdots & 0 & -g_1 \\
      -\beta_1 & d & -1 & 0 &0 & \cdots&0& 0 \\
      0 & -1 &  d & -1 & 0 &\cdots&0 &  0\\
      0 & 0 &  -1 & d & -1 &\cdots &0&  0\\
      0 & 0 & 0 & -1 & d & \cdots &0& 0\\
      \cdots& \cdots &\cdots&\cdots &\cdots&\ddots&\cdots&\cdots\\
      0 &\cdots&\cdots &\cdots&\cdots&\cdots&\ddots&-1\\
      -g_\Lambda & 0 & 0 & 0 & 0\cdots&\cdots&-1& d\\
    \end{array}\right)}
        \end{eqnarray}
which is our desired matrix if we set $d_1=d=2+\mu^2$, $\alpha_1=1$, $\beta_1=1$ , $g_1=g\otimes g^{-1}$ and $g_\Lambda=g^{-1}\otimes g$. We will evaluate to determinant for arbitrary
commuting parameters $d_i$,$\alpha_1$,$\beta_1$, $g_1$ and $g_\Lambda$.
Expanding the determinant in the first row 
\begin{equation}\label{DetMLambdaExpansion}
  {\rm Det} M_{\Lambda}=d_1{\rm Det} M_{\Lambda-1,0}-\alpha_1\beta_1{\rm Det} M_{\Lambda-2,0}-g_1\beta_1-g_\Lambda\alpha_1-g_1.g_{\Lambda}{\rm Det} M_{\Lambda-2,0} 
\end{equation}
where we have used the notation $M_{\Lambda,0}$ for the  tri-diagonal $\Lambda\times \Lambda$ matrix with $-1$ on the off diagonals and $d$ on the diagonal, i.e.  $M_{\Lambda}$ with $g_1$ and $g_{\Lambda}$ mapped to zero, which physically $M_{\Lambda,0}$ corresponds to imposing Dirichlet boundary conditions on the lattice Laplacian. Also we denote by $M_{\Lambda,1}$ the periodic tri-diagonal matrix $g_1$ and $g_\Lambda$ mapped to $1$ and with $-1$ on the
off diagonals and $d$ on the diagonal.

The matrix $M_{\Lambda,g}=a^2\Delta_{\Lambda,g}+\mu^2$ is obtained from (\ref{MLambda}) by setting $d_1=d=2+\mu^2$, $\alpha=\beta=1$, $g_{1}=g\otimes g^{-1}$ and $g_\Lambda=g\otimes g^{-1}$ in which case (\ref{DetMLambdaExpansion}) becomes
\begin{equation}\label{DetMLambda_g_Expansion}
  {\rm Det} M_{\Lambda,g}=(2+\mu^2){\rm Det} M_{\Lambda-1,0}-2{\rm Det} M_{\Lambda-2,0}-g\otimes g^{-1}-g^{-1}\otimes g
\end{equation}
We have already met ${\rm Det}M_{\Lambda,1}$ in  (\ref{M_Lambda_1}) and evaluated it in (\ref{DetMLamda1Eval}) finding that it is given by
\begin{equation}\label{DetM_Lam_1}
  {\rm Det}M_{\Lambda,1}=(2+\mu^2){\rm Det} M_{\Lambda-1,0}-2{\rm Det} M_{\Lambda-2,0}-2=z_{+}^\Lambda+z_{-}^{\Lambda}-2
\end{equation}
we can therefore conclude that\footnote{The determinant of
$M_{\Lambda,g}$ could equally be evaluated by the eigenvalue method,
however that would involve taking the $\Lambda$th root of $g$. We
avoid this complication by a direct evaluation expanding in
minors. This allows us to easily extend the analysis to discrete
groups.}
\begin{equation}
  {\rm Det} M_{\Lambda,g}=z_{+}^\Lambda+z_{-}^{\Lambda}-g\otimes g^{-1}-g^{-1}\otimes g=z_{+}^{\Lambda}({\bf 1}-z_{-}^{\Lambda}g\otimes g^{-1} )({\bf 1}-z_{-}^{\Lambda}g^{-1}\otimes g) 
\end{equation}
and therefore taking the remaining determinant over the group variables we have
\begin{equation}
{\rm\bf Det}M_{\Lambda,g}=z_{+}^{N^2\Lambda}{\rm\bf det}[{\bf 1}-z_{-}^{\Lambda}g\otimes g^{-1}]{\rm\bf det}[{\bf 1}-z_{-}^{\Lambda}g^{-1}\otimes g] 
\end{equation}
where we have used that the dimension of ${\bf 1}$ is $N^2$ --- this corresponds to the number of oscillators.
As discussed in the text above, taking the $N\rightarrow\infty$ of the inverse of the square root of the determinant and summing over $g$ and yields the Molien-Weyl formula
\begin{equation}
  Z_N(x)=\frac{1}{\vert G\vert}\sum_{g\in G}\frac{1}{{\rm\bf det}[{\bf 1}-x g\otimes g^{-1}]}\, .
\end{equation}

More generally if we take $g_1=R(g)$ and $g_\Lambda=R(g)^{-1}$ with $R$ any representation of the transformation group we obtain
\begin{equation}
{\rm\bf Det}M_{\Lambda,R}=z_{+}^{\nu\Lambda}{\rm\bf det}[{\bf 1}-z_{-}^{\Lambda}R(g)]{\rm\bf det}[{\bf 1}-z_{-}^{\Lambda}R(g)^{-1}] 
\end{equation}
where we denote the dimension of ${\bf 1}$ by $\nu$. 

Taking the limit of infinite $\Lambda$ and noting that limit $\displaystyle\lim_{\Lambda\rightarrow\infty}z_{\pm}^\Lambda\rightarrow =x^{\mp 1}$
yields the general result:
\begin{equation}
{\rm\bf Det} M_{\nu,R}=x^{\nu}{\rm\bf det}[{\bf 1}-x R(g)]{\rm\bf det}[{\bf 1}-x R(g)^{-1}] 
\end{equation}
{\bf Note:} This derivation has made minimal assumptions as to the nature or the representation, $R$, or the group. The Molien-Weyl result is then
obtained by summing over the group of the relevant power of the determinant. For generic non-hermitian matrices transforming as $X\rightarrow R\circ X$
one obtains
\begin{equation}
Z_\nu(x)=\frac{1}{\vert G\vert}\sum_{g\in G}\frac{1}{{\rm\bf det}[{\bf 1}-x R(g)]{\rm\bf det}[{\bf 1}-x R(g)^{-1}]}
\end{equation}
while in the real hermitian case we recover the familiar Molien-Weyl formula
\begin{equation}
  Z_\nu(x)=\frac{1}{\vert G\vert}\sum_{g\in G}\frac{1}{{\rm\bf det}[{\bf 1}-x R(g)]}
\end{equation}
with $R$ a real representation.

\section{Matrices with Dirichlet Boundary Conditions}\label{DirechletAppendix}
As an aside one can proceed to evaluate the Dirichlet determinants (i.e.  ${\rm Det}M_{\Lambda,0}$) directly as in the periodic case by the eigenvalue method, but it is slightly more involved. One can show that for even $\Lambda$ ($\Lambda=2n$)
\begin{equation}
  {\rm Det} M_{2n,0}=1+\sum_{k=1}^{2n} (z_{-}^{2k}+z_{+}^{2k})
\end{equation}
and for odd $\Lambda$ ($\Lambda=2n+1$)
\begin{equation}
  {\rm Det} M_{2n+1,0}=\sum_{k=1}^{2n+1} (z_{-}^{2k}+z_{+}^{2k})
  \end{equation}

These results can be proven by induction. One notes $z_{\pm}=\frac{1}{2}(d\pm \sqrt{d^2-4})$ so that $d=z_{-}+z_{+}$ and $z_{-}=z_{+}^{-1}$
and therefore
\begin{eqnarray}\label{recursion_Relns}
  d {\rm Det M}_{2n-1}&=(z_{-}+z_{+})\sum_{k=1}^{2n+1} (z_{-}^{2k}+z_{+}^{2k})={\rm Det} M_{2n,0}+{\rm Det} M_{2n-2,0}\nonumber\\
\hbox{and}&\\
  d{\rm Det M}_{2n}&=(z_{-}+z_{+})(1+\sum_{k=1}^{2n} (z_{-}^{2k}+z_{+}^{2k}))={\rm Det} M_{2n+1,0}+{\rm Det} M_{2n-1,0}\nonumber
\end{eqnarray}
which are the recursion relations to be solved.

Performing the sums and simplifying (using $z_{+}=z_{-}^{-1}$)
we find
\begin{equation}
  {\rm Det M}_{\Lambda,0}=f(z_{+},\Lambda)=f(z_{-},\Lambda) \qquad \hbox{where}\qquad   f(x,\Lambda)=\frac{x^{\Lambda+1}-\frac{1}{x^{\Lambda+1}}}{x-\frac{1}{x}}
\end{equation}
Observing that
\begin{equation}
  (x+\frac{1}{x})f(x,\Lambda-1)-2f(x,\Lambda-2)=x^\Lambda+\frac{1}{x^\Lambda}
  \end{equation}
directly establishes (\ref{DetM_Lam_1}). Also  (\ref{recursion_Relns}) is equivalent to
\begin{equation}
  (x+\frac{1}{x})f(x,\Lambda-1)
  =f(x,\Lambda)+f(x,\Lambda-2)\, .
  \end{equation}

\end{appendix}

\end{document}